\input harvmac

\font\cmss=cmss10 \font\cmsss=cmss10 at 7pt
\def\IZ{\relax\ifmmode\mathchoice
{\hbox{\cmss Z\kern-.4em Z}}{\hbox{\cmss Z\kern-.4em Z}}
{\lower.9pt\hbox{\cmsss Z\kern-.4em Z}}
{\lower1.2pt\hbox{\cmsss Z\kern-.4em Z}}\else{\cmss Z\kern-.4em Z}\fi}

\def\a{\alpha} \def\b{\beta} \def\g{\gamma} 
\def\d{\delta}  \def\ee{\varepsilon} 
 \def\th{\theta}  
\def\k{\kappa} \def\la{\lambda}  \def\m{\mu} \def\n{\nu}
  \def\p{\pi}  \def\r{\rho}
\def\s{\sigma}   
 \def\f{\phi}   

\def\pa{\partial}   \def\half{{1\over
2}} 
\def\oa{${\cal O}(\a ')$}\def\oaa{${\cal O}(\a '^2)$}
\def\wac{Weyl anomaly coefficients}

\Title{
\vbox{\baselineskip12pt\hbox{MIT-CTP-2668}
\hbox{hep-th/9708110}}}
{Duality and the Renormalization Group\footnote{*}
{\baselineskip12pt This work is supported in part by
funds provided by
the U.S. Department of Energy (D.O.E.) under cooperative
research agreement \#DF-FC02-94ER40818, and by NSF Grant
PHY-92-06867. E-mail:
{\tt haagense@ctp.mit.edu}.} }\vskip-0.15in

\centerline{
\vbox{\hsize3in\centerline{Peter E. Haagensen}}}
{\it
\centerline{Center for Theoretical Physics}\vskip-.15cm
\centerline{Laboratory for Nuclear Science}\vskip-.15cm
\centerline{Massachusetts Institute of Technology}\vskip-.15cm
\centerline{77 Massachusetts Avenue}\vskip-.15cm
\centerline{Cambridge, MA 02139, USA}}\vskip2cm

\nref\damgaard{P.H.~Damgaard and P.E.~Haagensen,
{\tt cond-mat/9609242}, {\it J.~Phys.} {\bf A30} (1997) 4681.}
\nref\burgess{C.P.~Burgess and C.A.~L\"utken,
{\tt cond-mat/9611118}, to be published in {\it Nucl.~Phys.} {\bf B}.
See also C.A.~L\"utken, {\it Nucl.~Phys.} {\bf B396} (1993) 670 for 
earlier considerations of how modular symmetry may constrain RG flows.}
\nref\haagensena{P.E.~Haagensen, {\it Phys.~Lett.} {\bf 382B}
(1996) 356.}
\nref\olsen{P.E.~Haagensen and K.~Olsen,
{\tt hep-th/9704157}, to be published in {\it Nucl.~Phys.} {\bf B}.}
\nref\schiappa{P.E.~Haagensen, K.~Olsen and R.~Schiappa,
{\tt hep-th/9705105}.}
\nref\penati{P.E.~Haagensen, S.~Penati, A.~Refolli and D.~Zanon, work in 
progress.}
\nref\buscher{T.H.~Buscher, {\it Phys.~Lett.} {\bf 194B}
(1987) 59; {\it Phys.~Lett.} {\bf 201B} (1988) 466.}
\nref\tseytlin{A.A.~Tseytlin, {\it Phys.~Lett.} {\bf 178B}
(1986) 34; {\it Nucl.~Phys.} {\bf B294} (1987) 383.}
\nref\balog{These were first written down explicitly in A.A.~Tseytlin, 
{\it Mod.~Phys.~Lett.} {\bf A6} (1991) 1721. Corrections to duality
transformations were also considered in J.~Panvel, {\it Phys.~Lett.} 
{\bf 284B} (1992) 50, and in
J.~Balog, P.~Forg\'acs, Z.~Horv\'ath and L.~Palla, {\it Nucl.~Phys.} 
(Proc.~Suppl.) {\bf B49} (1996) 16; {\it Phys.~Lett.} {\bf 388B} (1996) 121.
In the case of $D$ target isometries, duality at two loop order was also 
considered in K.~Meissner, {\it Phys.~Lett.} {\bf 392B} (1997) 298.}
\nref\jack{ I.~Jack and D.R.T.~Jones, {\it Nucl.~Phys.} {\bf B303} 
(1988) 260.}
\nref\meissner{N.~Kaloper and K.~Meissner, {\tt hep-th/9705193}.}
\nref\mavromatos{N.E.~Mavromatos and J.L.~Miramontes, {\it Phys.~Lett.} {\bf 201B}
(1988) 473.}
\nref\metsaev{R.R.~Metsaev and A.A.~Tseytlin, {\it Nucl.~Phys.} {\bf B293} 
(1987) 385.}
\nref\zanon{D.~Zanon, {\it Phys.~Lett.} {\bf 191B} (1987) 363.}

\bigskip

\centerline{\bf Abstract}\medskip

\vbox{\baselineskip12pt The requirement that duality and renormalization 
group transformations commute
as motions in the space of a theory has recently been explored
to extract information about the renormalization flows in 
different statistical and field theoretical systems. After a review
of what has been accomplished in the context of $2d$ sigma models, 
new results are presented which set up the stage for a fully generic
calculation at two-loop order, with particular emphasis on the question
of scheme dependence. }
\vfill

\vbox{\baselineskip12pt
\noindent{\it Talk presented at the NATO Workshop ``New Developments in 
Quantum Field Theory'', Zakopane, Poland, June 14 -- 20, 1997.}}
\Date{08/97}

\baselineskip14pt
\newsec{Introduction}\bigskip

Duality symmetries are typically transformations in the
parameter space of a theory which leave the partition function and the
correlators invariant (perhaps up to some known function of the parameters).
Renormalization group (RG) transformations also act in this same 
space, with similar invariance properties. Given the generality of this
observation, one expects that it may be possible to investigate the 
interplay between duality and the RG 
whenever a system presents a duality symmetry and a 
renormalization flow, regardless of whether it is a quantum field
theory, a statistical system, a lattice theory, etc..

Indeed, such a nontrivial interplay has recently been verified in a number
of different contexts \damgaard -\schiappa\ and a requirement
of consistency of duality symmetry and the RG  has been used to obtain
constraints on the RG flows. Spin systems, which generally
enjoy a Kramers-Wannier duality symmetry, were considered in \damgaard.
The Quantum Hall system, on the other hand, is strongly suspected to 
exhibit a much richer duality, under $SL(2,\IZ )$ or one of its 
(level 2) subgroups, and it was studied in \burgess. For the spin systems 
considered in \damgaard, the parameter space consisted of a single relevant
coupling (the inverse temperature), while Kramers-Wannier duality 
forms the (colloquially speaking) simple group $\IZ_2$, so that the
constraints on the RG structure end up not being strong enough to actually
determine unique flows. A more restrictive scenario appears in the
Quantum Hall system, where the parameter space consists of the upper complex
half-plane, and the requirements of holomorphy and modular symmetry naturally
turn out to be considerably richer. Yet, even in that case, the
existing results are not entirely conclusive: while on the one hand there
is not enough experimental data confirming the precise symmetry group of
the system, on the other hand by postulating a specific modular 
symmetry one still does not obtain unique RG flows.

Two-dimensional sigma models targeted on an arbitrary background of
metric, antisymmetric tensor and dilaton fields also present a duality
symmetry (when the background has an abelian isometry), and in that context, 
the situation is more favorable:
while the symmetry group is, like for spin systems, also $\IZ_2$, the
parameter space is of course much larger, and the action of the group on
it rather more involved, with geometry and torsion mixing in a nontrivial
way. For the loop orders and backgrounds considered so far, this has 
in fact allowed for an essentially complete determination of the RG flows
using only the requirement of consistency between duality and the RG
(the qualification `essentially' will be understood more clearly below). 
It is to these models and the relevant consistency requirements
that the bulk of what follows will be dedicated. 

To begin, we consider a system with a number of couplings, $k^i, i=1,\ldots 
n$, and a duality symmetry, $T$, such that
\eqn\oo{ Tk^i\equiv\tilde{k}^i=\tilde{k}^i(k)}
represents a map between equivalent points in the parameter space (with
equivalence taking the same meaning as, for instance, the order-disorder
equivalence in the $2d$ Ising model). We will also assume the system has
a renormalization group flow, $R$, encoded by a set of beta functions:
\eqn\pp{ Rk^i\equiv\b^i(k)=\m{d\over d\m}k^i\ ,}
with $\m$ some appropriate subtraction scale. On a generic function  
in the parameter space, $F(k)$, these operations act as follows:
\eqn\rtf{\eqalign{ TF(k)&=\ F(\tilde{k}(k))\phantom{xxx} \cr
RF(k)&=\ {\d F(k)\over\d k^i}\cdot \b^i(k)\ .}}
For a finite number of couplings the derivatives above should be understood
as ordinary derivatives, whereas in the case of the sigma model these
will be functional derivatives, and the dot will imply an integration over 
spacetime.
The consistency requirement governing the interplay of duality and the RG
can now be stated very simply:
\eqn\zz{ [T,R]=0}
or, in words, that duality transformations and RG flows commute as motions
in the parameter space of the theory. This is the main concept to be 
explored, and from which most results will follow.  
It is easy to see that the above amounts to the following
consistency conditions:
\eqn\cc{\b^i(\tilde{k})={\d \tilde{k}^i\over\d k^j}\cdot 
\b^j(k)\ ,}
that is, under duality transformations the beta function must transform
as a ``form-invariant contravariant vector'' (to avoid confusion: we are 
borrowing the language of General Relativity here, but of course duality 
transformations have nothing to do with diffeomorphisms!). It is this 
``form-invariance'', {\it i.e.}, the fact that the functional form of $\b^i$ 
on the l.h.s.~above must be the same as the one on the r.h.s., that is mostly 
responsible for the severity of the constraints engendered.

For the $2d$ Ising model on a square lattice this yields a constraint which
is nontrivially satisfied by the (known) beta function of the model, although
it does not determine uniquely this beta function. In the Quantum Hall
system, on the other hand, the resulting constraint is that 
the beta function transform as a weight $-2$ modular form (strictly
speaking, negative weight modular forms do not exist, and this obstruction
is then circumvented by slightly relaxing the condition of holomorphy, but
these details will not concern us here). 

In what follows, we will explore
in detail the analogous constraints in the context of $2d$ bosonic sigma
models, in order of increasing complexity: Sections 2 and 3 contain a 
review of previously published work \haagensena ,\olsen ,\schiappa\ on, 
respectively,
the fully generic one-loop case and the purely metric two-loop case,
while Section 4 comprises results obtained in the course of more recent 
investigations \penati, and presents the setup for the fully generic
two-loop case, in the presence of torsion. In this case, 
where all possible backgrounds are included, the issue of scheme dependence
will also be discussed in some detail, as it arises for the first time
to complicate matters in a considerable way.

\newsec{Sigma Models at One-Loop Order}\medskip

Our starting point is the $d\!=\!2$ bosonic sigma model on a generic
$D\!+\! 1$-dimensional background $\{ g_{\m\n}(X),b_{\m\n}(X)\}$ of
metric and antisymmetric tensor, respectively, where $\m,\n =0,1,\ldots
,D=0,i$, so that the $\m\!  =\! 0$ component is singled out.  We shall
assume this sigma model has an abelian isometry, which will enable
duality transformations, and we shall consider the background above in
the adapted coordinates, in which the abelian isometry is made manifest
through independence of the background on the coordinate $\theta\equiv
X^0$.  The original sigma model action reads:
\eqn\kmkm{\eqalign{ S={1\over 4\p\a '}\int d^2\!\s\, &\left[
g_{00}(X) \pa_\a \th\pa^\a\th +2g_{0i}(X)\pa_\a\th\pa^\a
X^i+g_{ij}(X)\pa_\a X^i\pa^\a X^j \right. \cr &+\left. i\ee^{\a\b}
\left( 2b_{0i}(X)\pa_\a\th\pa_\b X^i+b_{ij}(X)\pa_\a X^i \pa_\b
X^j\right)\right]\, .}}
Throughout, all background tensors can depend
only on target coordinates $X^i$, $i=1,\ldots ,D$, and not on $\th$.

The duality transformations in this model are also well-known \buscher :
\eqn\duality{\eqalign{\tilde{g}_{00} =&\ {1\over g_{00}}\ \ ,\ \ 
\tilde{g}_{0i}={b_{0i}\over g_{00}}\ \ ,\ \  
\tilde{b}_{0i}={g_{0i}\over g_{00}}\ , \cr \tilde{g}_{ij} =&\
g_{ij}-{g_{0i}g_{0j}-b_{0i}b_{0j}\over g_{00}}\ \ ,\ \ 
\tilde{b}_{ij}=b_{ij}-{g_{0i}b_{0j}-b_{0i}g_{0j}\over g_{00}}\ .}}

The statement of classical duality is that the model defined on the
dual background $\{\tilde{g}_{\m\n},\tilde{b}_{\m\n}\}$ is simply a
different parametrization of the same model, given that the
manipulations used to derive the transformations essentially only
involve performing trivial integrations in a different order starting
from the path-integral in which the abelian isometry is gauged.

On a curved worldsheet, another background must be introduced,
that of the dilaton $\f (X)$, coupling to the worldsheet curvature
scalar.
The RG flow of background couplings is given by their respective beta
functions:
\eqn\betas{
\b^{g}_{\m\n}\equiv\m{d\over d\m}g_{\m\n}\ ,\;
\b^{b}_{\m\n}\equiv\m{d\over d\m}b_{\m\n}\ ,\;
\b^{\f}\equiv\m {d\over d\m}\f\ ,
}
while the trace of the stress energy tensor is found from the
Weyl anomaly coefficients \tseytlin
\eqn\weyl{\eqalign{
\bar{\beta}_{\mu\nu}^g =&\beta_{\mu\nu}^g+2\alpha '\nabla_\mu
\partial_\nu\phi+\nabla_{(\mu} W_{\n )}
\ , \cr \bar{\beta}_{\mu\nu}^b =&\beta_{\mu\nu}^b
+\alpha '{H_{\mu\nu}}^\lambda
\partial_\lambda\phi +{H_{\mu\nu}}^\lambda W_{\la}+\nabla_{[\mu} L_{\n ]}
\ ,\cr \bar{\beta}^\phi =&\beta^\phi
+\alpha '(\partial_\mu\phi)^2+\nabla^\mu\f W_{\m}
\ , 
}}
where $W_{\m}$ and $L_{\m}$ are some specific target vectors depending
on $g_{\m\n}$ and $b_{\m\n}$, and $(\m\n )=\m\n +\n\m , [\m\n ]=\m\n -\n\m$.
For the loop orders and backgrounds considered in Sections 2 and 3,
$W_{\m}=L_{\m}=0$, and we will henceforth disregard them.
Both the beta functions and the Weyl anomaly coefficients turn out to satisfy 
the consistency conditions, \cc . However, while the latter 
satisfy them exactly, the former satisfy them up to a target 
reparametrization \haagensena,\olsen. Since both
encode essentially the same RG information, for simplicity we will
consider RG motions as generated by the Weyl anomaly coefficients
in what follows. Thus, in the present context, the couplings are denoted by
$k^i = \{ g_{\m\n}, \b_{\m\n}, \f\}$, with $i={g,b,\f}$ labeling metric, 
antisymmetric tensor and dilaton backgrounds, and
our $R$ operation will in this case be defined, on a generic functional 
$F[g,b,\f]$ (and in principle at any loop order), to be
\eqn\RF{
RF[g,b,\f]={\d F\over\d g_{\m\n}}\cdot\bar{\b}_{\m\n}^g
+{\d F\over\d b_{\m\n}}\cdot\bar{\b}_{\m\n}^b
+{\d F\over\d \f}\cdot\bar{\b}^\f\ ,}
where the dot also indicates a spacetime integration. Duality 
transformations are given by
\eqn\TF{
TF[g,b,\f]=F[\tilde{g},\tilde{b},\tilde{\f}]\ ,}
where $\tilde{\f}$ will be defined shortly.

The consistency conditions to be satisfied, \cc, 
that obtain from \duality\ translate more explicitly into:
\eqn\consistency{\eqalign{
\bar{\b}^{\tilde{g}}_{00} =&-{1\over g_{00}^2}
\bar{\b}^g_{00}\ , \cr \bar{\b}^{\tilde{g}}_{0i} =&
-{1\over g_{00}^2}\left(
b_{0i}\bar{\b}^g_{00}-\bar{\b}^b_{0i}g_{00} \right)\ , \cr
\bar{\b}^{\tilde{b}}_{0i} =&-{1\over g_{00}^2}\left(
g_{0i}\bar{\b}^g_{00}-\bar{\b}^g_{0i}g_{00} \right)\ ,\cr
\bar{\b}^{\tilde{g}}_{ij} =&\bar{\b}^g_{ij}-{1\over g_{00}}\left(
\bar{\b}^g_{0i}g_{0j}+
\bar{\b}^g_{0j}g_{0i}-\bar{\b}^b_{0i}b_{0j}
-\bar{\b}^b_{0j}b_{0i}\right) \cr
 &+ {1\over
g_{00}^2}\left( g_{0i}g_{0j}-b_{0i}b_{0j}\right) \bar{\b}^g_{00}\ , \cr
\bar{\b}^{\tilde{b}}_{ij} =&\bar{\b}^b_{ij}-{1\over g_{00}}\left(
\bar{\b}^g_{0i}b_{0j}+
\bar{\b}^b_{0j}g_{0i}-\bar{\b}^g_{0j}b_{0i}
-\bar{\b}^b_{0i}g_{0j}\right) \cr
 &+ {1\over
g_{00}^2}\left( g_{0i}b_{0j}-b_{0i}g_{0j}\right) \bar{\b}^g_{00}\ , }}
where, in a condensed notation, we take the quantities on the l.h.s.~above
to mean $\bar{\b}^{\tilde{g}}_{\m\n}\equiv\bar{\b}^g_{\m\n}[
\tilde{g},\tilde{b},\tilde{\f}]$, etc..
Both the dilaton duality transformation and its attendant consistency
condition are still ostensibly missing, but will be determined shortly.

At loop order $\ell$, the possible tensor structures $T_{\m\n}$ appearing
in the beta function must scale as
$T_{\m\n}(\Lambda g,\Lambda b)=\Lambda^{1-\ell}T_{\m\n}(g,b)$
under global scalings of the background fields.
At \oa\ one may then have
\eqn\tensorone{\eqalign{
\b_{\m\n}^{g} =&\a'\left( A\, R_{\m\n}+B\, H_{\m\la\r}H_{\n}^{\; \la\r}
+C\, g_{\m\n}R
+D\, g_{\m\n}H_{\a\b\g}H^{\a\b\g}\right)\ , \cr
\b_{\m\n}^{b} =&\a'\left(E\, \nabla^{\la}H_{\m\n\la}\right)\ ,
}}
where $H_{\m\n\la}=\pa_\m b_{\n\la}+{\rm cyclic\ permutations}$ and
with $A,B,C,D,E$ being determined from one-loop Feynman diagrams. As found 
in \haagensena, requiring \consistency\ to be satisfied,
and choosing $A=1$ determines $B=-1/4$, $E=-1/2$, and $C=D=0$,
independently of any diagram calculations. The consistency conditions,
\consistency, on
$g_{\m\n}$ and $b_{\m\n}$ alone also
allow for an independent determination of the dilaton
transformation (or ``shift'') $\tilde{\f}=
\f-{1\over2}\ln g_{00}$. From this shift, and \cc, one obtains
yet another consistency condition, 
\eqn\qwer{
\bar{\beta}^{\tilde{\f}}=\bar{\beta}^\f -{1\over 2g_{00}}\bar{\beta}^g_{00}
\ ,}
from which one can finally find the dilaton beta function, thus 
completely determining all beta functions at this order:
\eqn\betasone{\eqalign{ \b_{\m\n}^g =&\a '\left( R_{\m\n} -{1\over4}
H_{\m\la\r} H_{\n}^{\ \la\r}\right) \cr 
\b_{\m\n}^b =&-{\a '\over2}\nabla_\la H^\la_{\ \m\n}\cr 
\b^\f  =& C-{\a '\over2}\nabla^2\f\ -{\a '\over24}\
H_{\m\n\la}H^{\m\n\la}\ , }}
up to the constant $C$.

\newsec{Two-Loop Order with Purely Metric Backgrounds }\medskip

At the next order $R$ is modified by the two-loop beta functions, and
one must determine the appropriate modifications in $T$ such that
$[T,R]=0$ continues to hold. We begin by working with restricted
backgrounds of the form
\eqn\metricrest{
g_{\mu\nu}=\pmatrix{a & 0\cr 0 &\bar{g}_{ij}}\ ,
}
and $b_{\m\n}\!=\!0$, so that no torsion appears in the dual background
either. It is useful to define at this point the following two
quantities:
$a_{i}\equiv\pa_{i}\ln a$, and $q_{ij}\equiv\bar{\nabla}_ia_j+\half a_ia_j$,
where barred quantities here and below refer to the metric $\bar{g}_{ij}$
(also, indices $i,j,\ldots$, are contracted with the metric $\bar{g}_{ij}$).
Within this class of backgrounds classical duality transformations
reduce to the operation $a\to 1/a$, and it is simple to determine the
possible corrections to $T$ from a few basic requirements, spelled out in 
detail in \schiappa. For conciseness, we will directly present the final
result for the corrected duality transformations \balog :
\eqn\tlduality{\eqalign{
\ln\tilde{a} =&-\ln a+\la\a' a_{i}a^{i}\ , \cr
\tilde{g}_{ij} =&g_{ij}=\bar{g}_{ij}\ ,\cr
\tilde{\phi} =&\phi -{1\over 2}\ln a +{\la\over4}\a' a_{i}a^{i}\ ,
 }}
where $\la$ is a constant that cannot be determined from the basic requirements. 
The consistency conditions that follow from the above are:
\eqn\tlconsistency{\eqalign{
{1\over\tilde{a}}\ \bar{\beta}^{\tilde{g}}_{00}  =&
-{1\over a}\bar{\beta}^g_{00}+2\la\alpha'\left[
a^{i}\partial_{i}\left( {1\over a}\bar{\beta}^g_{00}\right)
-{1\over2} a^{i}a^{j}\bar{\beta}^g_{ij}\right]\ , \cr
\bar{\beta}^{\tilde{g}}_{ij} =& \bar{\beta}^g_{ij}\ ,\cr
\bar{\beta}^{\tilde{\phi}} =& \bar{\beta}^\phi-{1\over 2a}
\bar{\beta}^g_{00}+{\la\over 2}\alpha'\left[ a^{i}\partial_{i}\left(
{1\over a}\bar{\beta}^g_{00}\right)
-{1\over2} a^{i}a^{j}\bar{\beta}^g_{ij}\right]\ . 
}}
The terms scaling correctly under $g\rightarrow \Lambda g$ at this order,
and thus possibly present in the beta function, are
\eqn\tltensors{\eqalign{
\b_{\m\n}^{g(2)}  =& A_1\, \nabla_\mu\nabla_\nu R
+A_2\,\nabla^2 R_{\mu\nu}+
A_3\, R_{\mu\alpha\nu\beta}R^{\alpha\beta} 
+A_4\, R_{\mu\alpha\beta\gamma}{R_\nu}^{\alpha\beta\gamma}
+A_5\, R_{\mu\alpha}{R_\nu}^\alpha \cr 
+A_6&\, R_{\mu\nu}R
+A_7\, g_{\mu\nu}\nabla^2 R +A_8\, g_{\mu\nu}R^2 +
A_9\, g_{\mu\nu}R_{\alpha\beta}R^{\alpha\beta}+A_{10}\,g_{\mu\nu}
R_{\alpha\beta\gamma\delta}R^{\alpha\beta\gamma\delta}
}}
(we have used Bianchi identities to reduce from a larger set of tensor
structures). 

It will suffice in fact to study the consistency conditions
for the $(ij)$ components, $\bar{\b}^{\tilde{g}}_{ij}=\bar{\b}^g_{ij}$, 
in order to determine the only structure satisfying all the consistency
conditions.

We write
\eqn\poipoi{
\bar{\b}^g_{ij}=\a'\left(\b_{ij}^{g(1)}+2\bar{\nabla}_{i}\pa_{j}\f\right)
+\a'^2\b_{ij}^{g(2)}\ ,
}
where $\b_{ij}^{g(1)}\!=\!R_{ij}\!=\!\bar{R}_{ij}-{1\over 2}q_{ij}$
is the one-loop beta
function, and perform the duality transformation \tlduality,
keeping terms to ${\cal O}(\a'^2)$. Using the fact that the
one-loop Weyl anomaly coefficient satisfies the one-loop consistency
conditions \consistency, we arrive at
\eqn\betatwo{
{\b}_{ij}^{\tilde{g}(2)}=\b_{ij}^{g(2)}-{1\over 4}\la a_{(i}\pa_{j)}
(a^ka_k)\ ,
}
where the duality transformation of $\b_{ij}^{g(2)}$ is given simply
by $a \rightarrow 1/a$ without $\a'$ corrections, since this is already
${\cal O}(\a'^2)$. Separating the possible tensor
structures into even and odd tensors under $a \rightarrow 1/a$,
\eqn\oiu{ \b_{ij}^{g(2)}=E_{ij}+O_{ij}\ , \; \; \;
\tilde{E}_{ij}=E_{ij}\ , \; \; \; \tilde{O}_{ij}=-O_{ij}\ ,}
the even structures drop out of \betatwo\ and we are left with
\eqn\odd{
O_{ij}={1\over 8}\la a_{(i}\pa_{j)}(a^ka_k)\ .}
We now perform a standard Kaluza-Klein reduction on the ten terms in
\tltensors\ to identify which if any satisfy this condition.
The results can be obtained using the formulas in the Appendix of
\olsen, and are as follows:
\eqn\kaluza{\eqalign{
 (1)&:\ \nabla_{i}\nabla_{j}R = \bar{\nabla}_{i}\bar{\nabla}_{j}
(\bar{R}-{q_ n}^{n})\ , \cr
 (2)&:\ \nabla^{2}R_{ij} = (\bar{\nabla}^{2}
+{1\over 2}a_{k}\bar{\nabla}^{k})
(\bar{R}_{ij}-{1\over 2}q_{ij})
-{1\over 4}a_{i}a_{j}{q_ n}^{n} \cr
 &\phantom{18mm}-{1\over 4}a^{k}a_{(i}\left(\bar{R}_{j)k}-{1\over 2}q_{j)k}
\right)\ , \cr
 (3)&:\ R_{i\alpha j\beta}R^{\alpha\beta} = {1\over 4}q_{ij}{q_ n}^{n}
+\bar{R}_{injm}(\bar{R}^{nm}-{1\over 2}q^{nm})\ , \cr
 (4)&:\ R_{i\alpha\beta\gamma}{R_ j}^{\alpha\beta\gamma} =
{1\over 2}q_{ik}{q_ j}^{k}+\bar{R}_{iknm}\bar{R_ j}^{knm}\ , \cr
 (5)&:\ R_{i\alpha}{R_j}^\alpha =
\bar{R}_{ik}{\bar{R}_j}^{\;\, k}-{1 \over 2}\bar{R}_{k(i}{q_{j)}}^k+
{1 \over 4}q_{ik}{q_j}^k\ ,\cr
 (6)&:\ R_{ij}R = (\bar{R}_{ij}-{1 \over 2}q_{ij})(\bar{R}-{q_n}^n)\ ,
 \cr
 (7)&:\ g_{ij}\nabla^{2}R = \bar{g}_{ij}\left[ \half a^{k}\partial_{k}
(\bar{R}-{q_ m}^{m})\right. \cr
 &\phantom{21mm}\left. \phantom{ {a\over b}}
+\bar{\nabla}^{k}\partial_{k}(\bar{R}-{q_ m}^{m})
\right]\ , \cr
 (8)&:\ g_{ij}R^{2} = \bar{g}_{ij}\left(\bar{R}-{q_ m}^{m}\right)^{2}\ ,
 \cr
 (9)&:\ g_{ij}R_{\alpha\beta}R^{\alpha\beta} = \bar{g}_{ij}
\left[ {1\over 4}({q_ m}^{m})^{2}+(\bar{R}_{km}-{1\over 2}q_{km})^{2}
\right]\ , \cr
 (10)&:\ g_{ij}R_{\alpha\beta\gamma\delta}R^{\alpha\beta\gamma\delta}
 =  \bar{g}_{ij}\left[ q_{km}q^{km}+\bar{R}_{k\ell mn}
\bar{R}^{k\ell mn}\right]\ . }}   
The respective odd parts are
$$\eqalign{
O_{ij}^{(1)} =&-\bar{\nabla}_{i}\bar{\nabla}_{j}
\bar{\nabla}_{n}a^{n}\ , \cr
O_{ij}^{(2)} =&{1\over 2}a_{k}\bar{\nabla}^{k}\bar{R}_{ij}
-{1\over 2}\bar{\nabla}^{2}\bar{\nabla}_{i}a_{j}
-{1\over 4}a_{i}a_{j}\bar{\nabla}_{k}a^{k}\ , \cr
O_{ij}^{(3)} =&-{1\over 2}\bar{R}_{injm}\bar{\nabla}^{n}a^{m}
+{1\over 8}a_{n}a^{n}\bar{\nabla}_{i}a_{j}
+{1\over 8}a_{i}a_{j}\bar{\nabla}_{n}a^{n}\ , \cr
O_{ij}^{(4)} =&{1\over 4}a_{k}a_{(i}\bar{\nabla}_{j)}a^{k}\ ,}$$

\eqn\oddtwo{\eqalign{
O_{ij}^{(5)} =&-{1 \over 2}\bar{R}_{k(i}\bar{\nabla}_{j)}a^k+
{1\over 8}a_{k}a_{(i}\bar{\nabla}_{j)}a^{k}\ ,\cr
O_{ij}^{(6)} =&-{1\over2}\bar{R}\bar{\nabla}_ia_j-\bar{R}_{ij}
\bar{\nabla}_na^n+{1\over4}a_ia_j\bar{\nabla}_na^n \cr
 &+{1\over4}a_na^n
\bar{\nabla}_ia_j\ ,  \cr
O_{ij}^{(7)} =&\bar{g}_{ij}\left[ {1\over 2}a^{k}\partial_{k}
(\bar{R}-{1\over 2}a_{m}a^{m})-\bar{\nabla}^{k}\partial_{k}
(\bar{\nabla}_{m}a^{m})\right]\ , \cr
O_{ij}^{(8)} =& \bar{g}_{ij}\left[-2(\bar{\nabla}^{k}a_{k})\bar{R}
+(\bar{\nabla}^{k}a_{k})a^{m}a_{m}\right]\ , \cr
O_{ij}^{(9)} =&\bar{g}_{ij}\left[ {1\over 4}(\bar{\nabla}^{k}a_{k})a^{m}a_{m}
-(\bar{\nabla}_{k}a_{m})\bar{R}^{km}\right. \cr
 &\; \; \; \; \;\left. +{1\over 4}
(\bar{\nabla}_{k}a_{m})a^{k}a^{m}
\right]\ , \cr
O_{ij}^{(10)} =& \bar{g}_{ij}
(\bar{\nabla}_{k}a_{m})a^{k}a^{m}\ . }} 

It is fortunate that none of these tensors contain purely
even structures, since such structures are left unconstrained (and thus
undetermined) by duality. The only odd term of the form
\odd\ comes from $A_{4}R_{\m\alpha\beta\gamma}
{R_\n}^{\alpha\beta\gamma}$, and
a detailed inspection shows that no linear combination of the
other terms gives rise to odd tensors generically of the form \odd. This
determines that, with the requirement of covariance of duality under
the RG, the \oaa\ term in the beta function is
\eqn\xxx{
\beta^{g(2)}_{\mu\nu}=\lambda R_{\mu\alpha\beta\gamma}{R_\nu}^
{\alpha\beta\gamma}\ .
}
One should now check that the corresponding (00) component also
satisfies its consistency condition.
A straightforward computation shows that it does,
and the determination of the two-loop beta function is thus complete.

Although we treated a restricted class of metric
backgrounds, the final result is valid for a generic metric, since none of
the possible tensor structures are built out of the off-block-diagonal
$g_{0i}$ elements alone (in which case our consistency conditions would
be blind to them, just as they are to the even terms $E_{ij}$).

Simply using the requirements that duality and the RG
commute as motions in the parameter space of the sigma model, we have
thus been able to determine the two-loop beta function to be
\eqn\iuy{
\beta_{\mu\nu}=\alpha' R_{\mu\nu}+{\alpha'}^2\lambda R_{\mu\alpha\beta
\gamma}{R_\nu}^{\alpha\beta\gamma}\ ,
}
for an entirely generic metric background, again without any Feynman diagram
calculations. Because we used an extremely restrictive class of
backgrounds, it was not possible to determine the value of $\lambda$
(the correct value is $\lambda={1\over2}$). However, we expect that,
similarly to what happens at ${\cal O}(\alpha')$, once a more generic
background is used in the consistency conditions, even this constant
should be determined. We now examine how to go about calculating in
such a generic background in an efficient way.

\newsec{Setup for the Fully Generic Two-Loop Case }\medskip

The inclusion of torsion at two-loop order brings with it a number 
of complications which one should try to minimize as much as 
possible. First, the number of new terms appearing in the beta
functions is greatly increased as compared to the purely metric case.
Moreover, there is now one more beta function to worry about, for the
antisymmetric tensor. Also, it is clear that there will be several
new terms in the corrections to duality transformations, and the
general arguments used in \schiappa\ will not be sufficient to determine them.
To finally complicate the situation further, the scheme dependence 
present leaves a lot more latitude to
what the correct expressions for these beta functions are, as well as 
which duality transformations should make them transform covariantly.

It thus seems that a direct guessing of the corrected duality 
transformations, attempting to keep the two-loop beta functions covariant,
would be an extremely arduous task, and we will try rather to first
streamline the calculations involved by going through what may seem
at first a longer path.

We start with the observation that there is a connection between the 
Weyl anomaly coefficients and the string background effective action
(``EA'' in what follows),
whereby one establishes a direct relation between the former and the
equations of motion of the latter. In principle, there is thus the 
possibility that the duality transformation properties of one imply
the transformation properties of the other. Should this be the case,
one might save considerable effort by studying the effective action
alone, since this is a scalar function on the parameter space, invariant
under duality, whereas the beta function represents a vector field in
that space, with nontrivial transformation properties. 

Unfortunately, we will see that the transformation properties of the 
EA under duality will {\it not} allow us to deduce the transformation
properties of the Weyl anomaly coefficients. However, the detailed 
consideration of this relationship will still be useful, on the one hand
to limit the possible transformations under which the Weyl anomaly
coefficients behave covariantly, and on the other, to clarify the 
messy issue of scheme dependence.

We begin by examining the situation at one-loop order. The EA is given
by
\eqn\ea{ S\equiv\a' S_0=\a'\int d\th\ d^D\! x\ \sqrt{g}e^{-2\f}\left( 
R+4(\nabla\f 
)^2 -{1\over12}H_{\m\n\la}H^{\m\n\la}\right)\ .}
Given the one-loop expressions for the beta functions, \betasone, 
it is simple to see that this EA is actually equivalent to (with $C=0$)
\eqn\srv{ S = R\ V=V_{,i}\cdot \bar{\b}^i\ ,}
where $V=\sqrt{g}\exp -2\f$, and $V_{,i}\equiv \d V/\d k^i$, in the
notation of the introduction. Because we know the one-loop \wac\ transform 
contravariantly under duality (cf. \cc ), the gradient 
$\d /\d k^i$
transforms covariantly, and $V$ is invariant, it immediately follows
that $S$ as defined above is invariant under duality transformations.
Similarly, at higher loop orders, if we are able to find the corrected 
duality transformations under which the higher-loop \wac\ transform
contravariantly as in \cc, and if we are able to find a scheme in
which the EA continues to be given by \srv, then we are 
guaranteed duality invariance of the EA.
But that is actually opposite to the direction we are seeking. Can we 
attempt to argue also conversely? At first sight, \srv\
does seem to give this converse result, that once a duality transformation
can be found at some loop order that keeps $S$ invariant, that will
imply the sought for contravariance of the \wac , and thus
the statement that $[T,R]=0$.

That is not correct, however, for $V_{,i}$, which is given more 
explicitly by
\eqn\vi{
V_{,i}= \pmatrix{V_{,g}\cr V_{,b}\cr V_{,\f}
}= \pmatrix{ -\half V g^{\m\n}
\cr 0\cr -2V}\ ,}
(we are omitting a delta function coming from the functional derivative)
has an enormous amount of ``zero modes'', given by
\eqn\zi{
z^{i}= \pmatrix{z^{g}\cr z^{b}\cr z^{\f}
}= \pmatrix{ z^{g}_{\m\n}
\cr z^{b}_{\m\n}\cr {1\over4}g^{\m\n}z^{g}_{\m\n}
}\ ,}
with $z^{g}_{\m\n},z^{b}_{\m\n}$ arbitrary, so that $V_{,i}\ z^{i}=0$.
This implies that, at some loop order, if there is a set of duality
transformations that keeps $S$ as defined in \srv\ invariant,
then the \wac\ are seen to transform not as in \cc, but as 
\eqn\ztrsf{
\b^i(\tilde{k})+\tilde{z}^{i}(\tilde{k})={\d \tilde{k}^i\over\d k^j}\cdot 
\left( \b^j(k)+z^{j}(k)\right)\ ,}
with $z^i(k)$ and $\tilde{z}^i(\tilde{k})$ some specific vectors of the
form \zi, not necessarily zero.
Naturally, this does not represent any covariance property at all. 

Could some other reasoning rescue the possibility that the invariance of
the EA might imply the contravariance of the \wac ? 
For instance, one immediate
criticism that may be applied to the argument above is that one is not
sure {\it a priori} that the EA should really be given by \srv.
This brings us to the issue of scheme ambiguities.

An independent definition of the EA corresponds to the field theory
action that generates the massless sector of the (string) tree level
string S-matrix. That EA contains a large number of terms that are
ambiguous in that they can be modified
with field redefinitions of the EA, and it contains a smaller number of
terms that are invariant under field redefinitions, and thus unambiguous.
Field redefinitions also affect the beta functions, and stemming from 
their definition, \betas, it is simple to see that they must 
transform
under field redefinitions as contravariant vectors (now we {\it are}
talking about diffeomorphisms). Parenthetically, we note that a subset of
these field redefinitions correspond to what is typically referred to as
a change of subtraction scheme in the renormalization of the sigma model:
if, say, minimal subtraction corresponds to the subtraction of a 
divergent term $1/\ee\ T_{\m\n}$, then a different, nonminimal scheme
corresponds to the subtraction of a term $({\rm const.} +1/\ee )\ T_{\m\n}$,
which in turn is equivalent to a field redefinition by the term
${\rm (const.)}\ T_{\m\n}$. With such a notion of scheme ambiguity, it
can be seen that the two-loop beta function in the purely metric case
is actually scheme independent, a property which is lost when torsion
is included. More generally, however, because the sigma model is not
renormalizable in the usual sense, one is also allowed finite subtractions
of terms not originally present in the action, and these correspond 
to arbitrary field redefinitions. In order not to propagate
semantic confusion, we will refrain from using the standard (and more
restrictive)
notion of scheme ambiguity, and will always consider instead the full
generality of arbitrary field redefinitions, referring to different
redefinitions as different choices of scheme.

At any rate, we realize from the discussion above that there is an 
unambiguous and independent notion to the EA, and that to each different 
``realization'' of it, or choice of scheme, there corresponds also a 
choice of scheme for the beta functions. It is expected that in any scheme
there should be a relation between the equations of motion of the EA and
the \wac\ of the form
\eqn\sgb{  {\d S\over\d k^i}=G_{ij}\cdot\bar{\beta}^j\ ,}
with $G_{ij}$ invertible, in the sense that the equations of motion
imply the vanishing of the \wac , and vice-versa (an even stronger 
requirement
would be the positivity of $G_{ij}$, in order to connect the EA to a 
$c$-function, but that will not concern us here). Could this relation 
between the EA and $\bar{\beta}^i$'s allow us to deduce the contravariance 
of the latter from invariance of the former?

The notation certainly is very suggestive, with $G_{ij}$ naturally appearing
to change an object transforming contravariantly into an object transforming
covariantly. However, insofar as $G_{ij}$ itself has no independent
meaning,\footnote{$^\dagger$}
{\baselineskip12pt Again, in the context of a $c$-theorem it would, but
scheme dependence in the present context complicates matters too much
to allow one at this point to seriously conjecture $G_{ij}$ to be the 
Zamolodchikov metric. This may well turn out to be true eventually, in some 
scheme, but we shall simply not assume it here.}
but is {\it devised} to have the above equation satisfied (in the
sense that it just represents the particular linear combinations of
$\bar{\beta}^i$'s that yield the equations of motion of the EA),
the answer is again unfortunately negative, and it is well exemplified by
the situation at one loop order already. With the beta functions given
by \betasone, and the EA by \ea, it is simple to 
find that $G_{ij}$ will be
\eqn\gij{ \eqalign{
G_{ij}=& \pmatrix{G_{gg}^{\m\n\la\r}& G_{gb}^{\m\n\la\r} & 
G_{g\f}^{\m\n}\cr
G_{bg}^{\m\n\la\r}& G_{bb}^{\m\n\la\r} & G_{b\f}^{\m\n}\cr
G_{\f g}^{\la\r}& G_{\f b}^{\la\r} & G_{\f\f}}\cr &\phantom{XX}\cr
=& \sqrt{g}e^{-2\f}
 \pmatrix{\half g^{\m\n}
g^{\la\r} -\half g^{\m (\la}g^{\r )\n}& 0 & -2g^{\m\n}\cr 
0 & -\half g^{\m [\la}g^{\r ]\n} & 0\cr
-2g^{\la\r}& 0 & 8}\ .}}

We already know that both the one-loop \wac\ and EA transform
in the proper way, and thus one {\it must} find, if one were to  
check explicitly the 
transformation properties of the particular $G_{ij}$ given above, that
it transforms like a rank 2 form-invariant covariant tensor under duality.
If we did not know how the $\bar{\beta}^i$'s transformed, however, all we 
could tell from the invariance of $S$ is that $G_{ij}$ has to cancel 
whatever (possibly completely wrong) transformation property of 
$\bar{\beta}^j$, and yield the transformation rule for a covariant vector. 
Thus if, say, the antisymmetric
tensor Weyl anomaly coefficient were twice its correct value, \sgb\
would still hold if we multiplied $G_{bb}$ by $1/2$, and yet the ``new"
$\bar{\beta}^i$ (with the wrong coefficient of $\bar{\beta}^b$) would
certainly {\it not} satisfy \cc, and consequently $[T,R]=0$ would
also not be satisfied. Accordingly, the ``new'' $G_{ij}$ would also not
transform like a rank 2 covariant tensor. Another clear, and even more 
pertinent, example of this can be seen with $G_{ij}$ at two-loop order: 
if \sgb\ is expanded to \oaa , the r.h.s. will contain, at \oaa ,
a term given by the contraction of $G_{ij}$ at two-loop order with 
$\bar{\b}^j$ at one-loop order. Such an expansion is considered in \jack, 
and the authors note there that because $G_{ij}$ at two-loop order
has the same tensor structures as $\bar{\b}^j$ at one-loop order, whenever
a term appears containing (roughly speaking) the square of a one-loop 
$\bar{\b}^j$,
it becomes impossible to determine which piece belongs to $G_{ij}$, and
which to $\bar{\b}^j$. Of course, any choice other than the correct one will 
lead to a violation of $[T,R]=0$, even though \sgb\ is perfectly 
satisfied whichever way these terms are split up. Incidentally, the 
authors of \jack\ suggest the only way to resolve this indefinition in
$G_{ij}$ is by going one order higher; we would suggest instead that 
the present considerations involving duality will eventually resolve this
problem without the need to go to three-loop order.

So far, it has seemed that the invariance of the EA cannot really be of any
help in determining the covariance properties of the \wac . But, in fact, the
above has not been entirely in vain: we know that, in the scheme in which 
the EA is given by \srv\ at higher loop order, if there exists at all
any duality transformations that respect $[T,R]=0$, {\it i.e.}, such that 
$\bar{\beta}^i$ transforms contravariantly, then these transformations must
keep the EA invariant; thus, if there is only one set of transformations that
keep the EA invariant, these are the only transformations that have a chance
of satisfying $[T,R]=0$. So, we are not guaranteed that the transformations
that keep $S$ invariant satisfy $[T,R]=0$, but if we know the only 
transformations that keep $S$ invariant, we are at least not groping in 
the dark trying to guess which transformations we should be testing on 
the \wac . 

Recently, a set of corrected duality transformations has been found \meissner\
that keep invariant the two-loop EA in a particular scheme. In that scheme,
it is claimed that the matrix $G_{ij}$ connecting the equations of motion and
the \wac\ is purely numerical, that is, it contains no spacetime derivative
operators acting on $\bar{\beta}^i$ \jack ,\mavromatos .
That is certainly a crucial advantage
if one is interested in studying a $c$-theorem for generalized sigma models.
For our purposes, however, the disadvantage of that scheme is the fact that
the expression for the EA is very complicated, containing a large number
of scheme dependent terms as compared to the ``minimal" EA that reproduces
string scattering amplitudes. Furthermore, that EA does not satisfy
\srv, a property we would like to maintain; instead, the 
so-called ``minimal" EA, $S_{min}$, does \metsaev .
We would therefore like to obtain all our results in that scheme if 
possible.

In order to do this, we will study the general problem of scheme dependence,
to determine whether we can find a set of
duality transformations that keeps an EA invariant in one scheme
if another set of transformations is given that keeps the EA invariant 
in another scheme. 

In the generic notation of the introduction, we assume we are given
an EA in one particular scheme to two-loop order, 
\eqn\eaone{  S(k)=\a'S_0(k)+\a'^2\ S_1(k)\ ,}
and a set of (two-loop corrected) duality transformations
\eqn\done{  \tilde{k}^i(k)=\tilde{k}_0^i(k)+\a'\ \tilde{k}_1^i(k)\ ,}
such that $S(\tilde{k})=S(k)$ to \oaa . Thus,
$S_0(k)$ is given by \ea, $S_1(k)$ may be for instance the 
two-loop EA in the scheme considered in \jack,\meissner, 
$\tilde{k}_0(k)$ is given
by \duality, and $\tilde{k}_1(k)$ would then be the corrections
to duality found in \meissner. We now make a field redefinition
\eqn\redef{  \hat{k}^i(k)= k^i+\a'\ f^i(k)\ ,}
with $f^i(k)$ some functional of the couplings with the appropriate 
dimensions. The field redefined EA, to \oaa , will be
\eqn\earedef{  \hat{S}(k)\equiv S(\hat{k})=S(k)+\a'\ f^i(k)\cdot
{\d S(k)\over\d k^i}\ .}
To the order considered, the invariance $S(\tilde{k})=S(k)$ assumed 
implies
\eqn\ss{\eqalign{  S_0(k) =&S_0(\tilde{k}_0) \cr
S_1(k)  =&S_1(\tilde{k}_0)+\tilde{k}_1^i(k)\cdot
{\d S_0(\tilde{k}_0)\over\d \tilde{k}_0^i}\ .}}
We now write
\eqn\dredef{  \tilde{\k}^i(k)=
\tilde{k}_0^i(k)+\a'\ \tilde{\k}_1^i(k)}
for the duality transformations that will keep the field redefined EA,
$\hat{S}(k)$, invariant: 
\eqn\earedefinv{  \hat{S}(\tilde{\k}(k))=\hat{S}(k)\ .} 
To determine these field redefined duality transformations, one must 
now substitute \dredef\ into \earedefinv,
using \redef,\earedef,\ss, and
keeping terms to \oaa . This is done in a straightforward way, and we 
simply state the final result:
\eqn\kappaone{  \tilde{\k}_1^i(k)=\tilde{k}_1^i(k)+
\left( {\d\tilde{k}_0^i \over\d k^j}\cdot f^j(k)-f^i(\tilde{k}_0)
\right)\ .}
This is the result we sought: given a set of transformations keeping
the EA invariant in some scheme, we can explicitly construct the
set of transformations keeping the EA invariant in any other scheme.
It is interesting to note that the term in parenthesis on the r.h.s. above
represents precisely the one-loop consistency conditions $[T,R]=0$, but 
acting on the field redefinitions rather than on the \wac . In other words,
in changing from one scheme to another through a field redefinition, 
the duality transformations that keep the redefined EA invariant
differ from the original transformations by a term which ``corrects'' for
how much off the field redefinitions themselves are from satisfying
the one-loop consistency conditions.

The minimal EA, $S_{min}=\a'S_0+\a'^2\ S_{1min}$,
\eqn\smin{\eqalign{ 
S_{1min}=& {1\over4}\int d\th\ d^D\! x\ \sqrt{g}e^{-2\f}\left( 
R_{\m\n\la\r}R^{\m\n\la\r}-\half R^{\m\n\la\r}H_{\m\n\s}H_{\la\r}^{\ \ \s}
\right. \cr &+\left.
{1\over24}H_{\m\n\la}H^{\n}_{\ \r\a}H^{\r\s\la}H_{\s}^{\ \m\a}
-{1\over8}H_{\m\a\b}H_{\n}^{\ \a\b}H^{\m\r\s}H^{\n}_{\ \r\s}
\right) \ ,}}
the field redefinition taking the nonminimal action of \jack,\meissner\
into it, and
the beta functions in several different subtraction schemes, can all be 
gleaned from the literature \jack ,\metsaev ,\zanon . The task at hand 
is now to 
find the duality transformations that keep \smin\ invariant
and, using those as the operation $T$, and the beta functions in the
appropriate scheme to define $R$, verify whether $[T,R]=0$ 
holds at two-loop order. It should be noted that what we have done 
above {\it guarantees} that there exists a set of duality 
transformations that keeps $S_{min}$ invariant; however, the 
constructive procedure, in \kappaone,
of obtaining these transformations starting from 
the transformations found in \meissner, is very likely not the 
most efficient way
to proceed, and we have opted instead for  direct guessing and verification
on $S_{min}$. We expect to report on this in the near future \penati . 

\newsec{Conclusions}\medskip

The requirement that duality and the RG commute as motions in the parameter
space of a model is a very basic one, and it has been shown not only to be
verified in the instances it has been tested, but also to yield important
constraints on the RG flows in the context of $2d$ sigma models. 

While at one-loop order this interplay between duality and the RG in the
sigma model has been thoroughly investigated, at two-loop order the 
analysis has not been exhaustive so far. To help us in achieving this 
complete analysis, we have available
first of all a set of duality transformations
keeping a string background effective action invariant \meissner . 
We have shown that there is no guarantee that the set of transformations
that keeps this effective action invariant will also turn out to satisfy
the duality consistency conditions $[T,R]=0$. However, we have also seen
that if any transformations at all exist that do satisfy the consistency 
conditions,
they must also keep the effective action invariant (at least in the 
``minimal'' scheme), so that by finding
the transformations that keep the effective action invariant one is 
selecting the one set of transformations that has a chance of 
satisfying $[T,R]=0$. 

We believe this basic statement, $[T,R]=0$, to be a more 
fundamental feature of the models in question than the invariance of the 
string background effective action, which follows from it (and which 
only is defined for sigma models). This has represented sufficient 
motivation for us to delve into the question of its validity in full 
generality at two-loop order, with the encouragement that the
existence of a duality invariance
of the string background effective action has already been shown
in the same context.

Field redefinition ambiguities enter at this loop order as an added
complication. We have taken the first step in comprehensively
accounting for them by establishing that duality symmetry is a well-defined
notion over and above the presence of such ambiguities, in the sense that
if it is present in one choice of scheme, it may be modified but will
nonetheless also be present in any other scheme.  
\medskip

{\bf Acknowledgments}\medskip

It is a pleasure to thank the Workshop organizers, P.~H.~Damgaard, 
J.~Jurkiewicz and M.~Prasza\l owicz, for the invitation to present
this work.

\listrefs
\end